\documentclass{zermatt}
\usepackage{graphicx}
\usepackage{natbib}
\usepackage{mathrsfs,amssymb}
\usepackage[T1]{fontenc}
\usepackage[utf8]{inputenc}
\usepackage[english]{babel}
\usepackage[breaklinks, colorlinks, citecolor=blue]{hyperref}

\newcommand{\zzeta}{\zeta_{\rm H_2}^{\rm ion}}
\newcommand{\nhh}{N_{\rm H_2}}
\begin{document}
%-----------------------------
\title{On the origin of cosmic-ray ionisation in star-forming regions} 
%\runningtitle{Ossenkopf-Okada et al.: Instructions for proceedings articles}
%
\author{Marco Padovani}
\address{INAF-Osservatorio Astrofisico di Arcetri, Largo E. Fermi 5, 50125 Firenze, Italy\\ \email{marco.padovani@inaf.it}}
%\secondaddress{Astronomers-For-Planet-Earth Germany}
%\author{R.~Schaaf}
%\address{Argelander-Institut für Astronomie, Auf dem Hügel 71, 53121 Bonn, Germany, \email{rschaaf@astro.uni-bonn.de}}
%\author{A.~Stutz}
%\address{Departamento de Astronom\'{i}a, Universidad de Concepci\'{o}n,Casilla 160-C, Concepci\'{o}n, Chile,
%\email{astutz@astro-udec.cl}}
%\author{I.~Breloy}
%\sameaddress{1}
%
\begin{abstract}
A field with particularly exciting results over the past few years is the study of the interaction of cosmic rays with interstellar matter. 
For star formation to take place, gas and dust need to be sufficiently cold for gravity to overcome thermal pressure, 
and the ionisation fraction must be low enough to enable substantial decoupling between the gas and the Galactic magnetic field. 
As soon as the visual extinction is of the order of $3-4$ magnitudes, the ultraviolet photon flux from the interstellar radiation field is fully quenched, 
thus the only source of ionisation and heating is provided by low-energy cosmic rays.
We will briefly focus on the Galactic and local origin of cosmic rays and on their effects on medium ionisation. 
%Within this short article we give some hints how to write your proceedings article for the Chile-Cologne-Bonn-Symposium 2022. All aricles should be prepared using this template and compilation with \texttt{pdflatex} should be tested. We demonstrate a number of peculiar cases directly in the source code of the template.\\
%	The proceedings will be published as electronic book through the University of Cologne. Articles must be uploaded by \textbf{November 30, 2022} to https://uni-koeln.sciebo.de/s/ngmLXNGqbbTCvR6. Strict naming conventions apply.
%The size of the contributions is limited to at most 3 pages for posters and contributed talks, 5 pages for invited talks, and 7 pages for reviews.
\end{abstract}
\maketitle
%%-----------------------------
%%      your text
%%-----------------------------
\section{Tracing the presence of cosmic rays in star-forming regions}\label{Padovani:sec:intro}

Cosmic rays (CRs) have a major impact on the
chemical and dynamical evolution of star-forming regions.
Their energy density ($\sim1~$eV~cm$^{-3}$) is of the order of that of
the cosmic microwave background, the visible starlight, and the Galactic magnetic field
\citep{Ferriere2001}. This makes CRs a major contributor to the energy budget of the insterstellar medium.

Key questions such as 
$(i)$ what are the mechanisms that determine the collapse of a molecular cloud, 
$(ii)$ what are the basic processes that regulate the growth of dust grains, 
$(iii)$ what is at the origin of the chemical complexity observed in clouds, and 
$(iv)$ who is responsible for energetic phenomena such as the synchrotron emission observed at different scales of a 
cloud, find a common denominator in CRs \citep[for a review, see][]{Padovani+2020}.
More precisely, it is the low-energy CRs ($E<{\rm TeV}$) that are relevant in 
answering the above questions, because 
the cross sections of the main processes occurring in molecular clouds (e.g., ionisation, dissociation, and 
excitation of H$_2$ by CR protons and electrons) peak between about 10~eV and 10~keV 
\citep[see, e.g.][]{Padovani+2009,Padovani+2018a}.
The core parameter that is employed in astrochemical codes to interpret the observed molecular abundances, as 
well as in non-ideal magnetohydrodynamic codes used to simulate, e.g., the collapse of a molecular cloud,
is the so-called CR ionisation rate $\zzeta$. 
It represents the number of ionisations of 
hydrogen molecules by CRs per unit time and is defined by
\begin{equation}
\zzeta = 4\pi\int_I^\infty j(E)[1+\Phi(E)]\sigma^{\rm ion}(E)~{\rm d}E\,,
\label{Padovani:eq:1}
\end{equation}
where $I=15.44$~eV is the H$_2$ ionisation threshold, $j(E)$ is the CR flux (or spectrum), namely
the number of CRs per unit energy, time, area, and solid angle, the factor $1+\Phi(E)$ accounts for
further ionisation by secondary electrons \citep[see][]{Ivlev+2021}, and 
$\sigma^{\rm ion}(E)$ is the ionisation cross section.

Over time, numerous techniques have been developed to estimate $\zzeta$ through observations of molecular tracers. Depending on the region of the molecular cloud being observed, different tracers are used.
In diffuse regions of molecular clouds, one of the most reliable, due to its simple network of formation and destruction reactions, is H$_3^+$ \citep{Oka2006}, followed by
OH$^+$, H$_2$O$^+$ \citep{Neufeld+2010}, and ArH$^+$ \citep{NeufeldWolfire2017}.
In denser regions, other tracers are used, such as HCO$^+$, DCO$^+$, and CO in low-mass dense cores 
\citep{Caselli+1998,Redaelli+2021}, HCO$^+$, N$_2$H$^+$, HC$_3$N, HC$_5$N, and c-C$_3$H$_2$ in protostellar clusters 
\citep{Ceccarelli+2014,Fontani+2017,Favre+2018}, and, more recently, H$_2$D$^+$ and other H$_3^+$ isotopologues
in high-mass star-forming regions \citep{Bovino+2020,Sabatini+2020}, PO$^+$ towards the Galactic centre
\citep{Rivilla+2022}, and H$_3$O$^+$, SO, and HCN isomers in the central molecular zone of the near starburst galaxy
NGC~253 \citep{Holdship+2022,Behrens+2022}.

Each of these methods involves a different degree of uncertainty, which affects the degree of accuracy in determining 
the ionisation rate. For example, H$_3^+$ observations can be carried out only towards specific lines of sight in the direction of early-type background stars. Moreover, $\zzeta$ obtained through H$_3^+$ is proportional to the 
gas volume density, the local ionisation fraction, and the details of the interstellar ultraviolet field attenuation,
all of which affects the resulting ionisation rate estimate.
In denser regions, the main drawback is that chemistry is much more complex than in diffuse clouds. 
This requires up-to-date, extensive reaction networks, and the main uncertainties come from 
the destruction and formation rates of several species, which are not well known, 
as well as the unconstrained amount of carbon and oxygen depletion on dust grains.
Recently, \citet{Bialy2020} introduced a new method to estimate $\zzeta$ from 
the observations of near-infrared rovibrational transitions of H$_2$, 
mainly excited by secondary CR electrons. This method, refined and extended by \citet{Padovani+2022},
has been tested by \citet{Bialy+2022}, obtaining
upper limits of $\zzeta$. In principle, the
James Webb Space Telescope should be able to detect these near-infrared H$_2$ lines, 
making it possible to derive, for the first time, spatial variation of $\zzeta$ in dense gas and to
test competing models of CR propagation and attenuation in the interstellar medium
\citep{EverettZweibel2011,MorlinoGabici2015,SilsbeeIvlev2019,Padovani+2018b,Gaches+2021}.

\section{Galactic or local cosmic rays?}
\begin{figure}[htp]
      \begin{minipage}{0.60\textwidth}
	\includegraphics[angle=0.0,width=\textwidth]{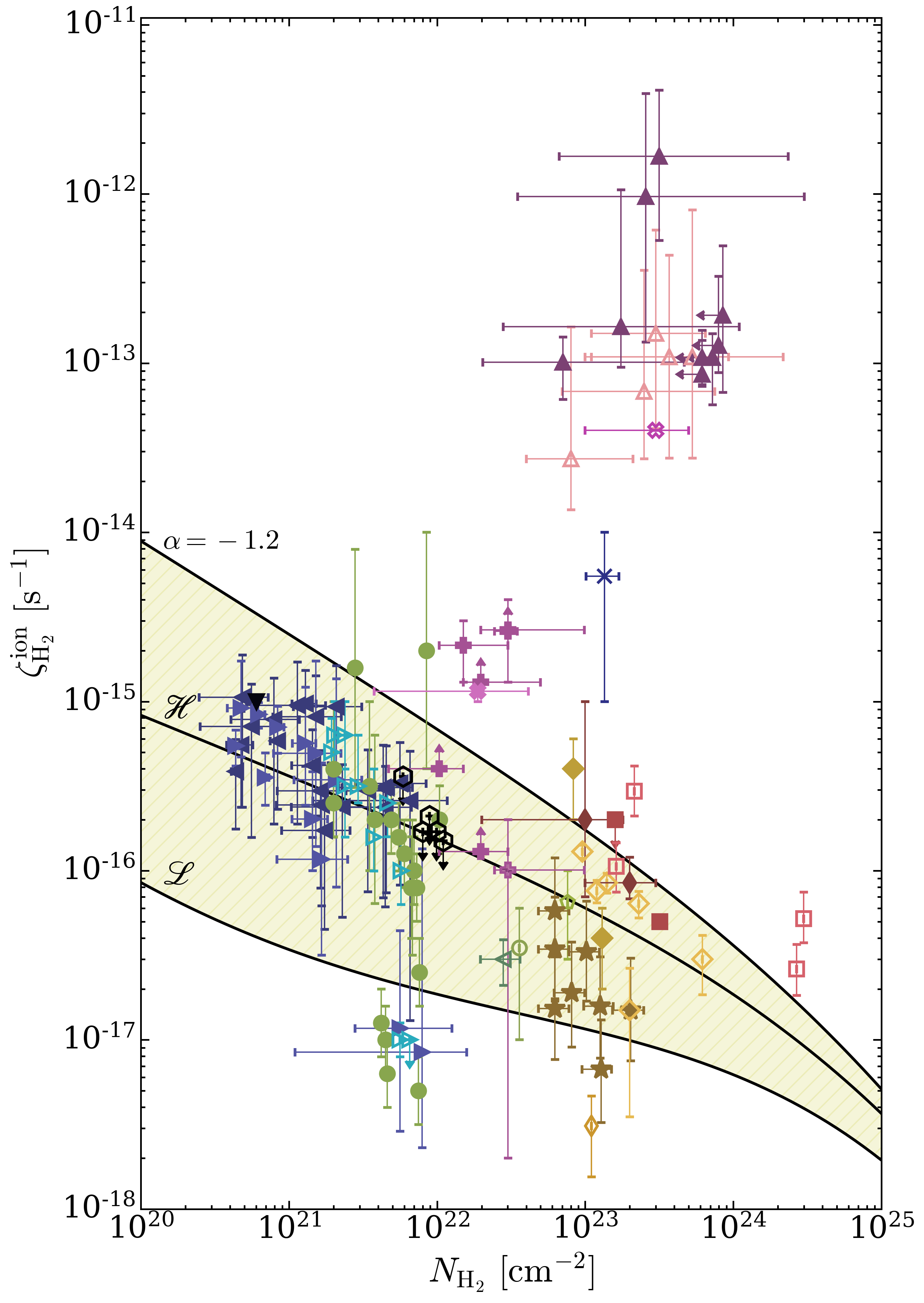}
      \end{minipage}
      \hspace{0.5cm}
      \begin{minipage}[!t]{0.35\textwidth}
	\caption{Total CR ionisation rate as a function of the H$_{2}$ column density: theoretical model $\mathscr{L}$,  $\mathscr{H}$
and with low-energy spectral slope $\alpha=-1.2$ (solid black lines) from \citet{Padovani+2022}.
The hatched-filled region show the expected $\zzeta$ range in case where ionisation is due to the average Galactic CR flux. 
Observational estimates: in diffuse clouds by
\citet[][solid downward-pointing triangle]{Shaw+2008},
\citet[][solid left-pointing triangles]{IndrioloMcCall2012},
\citet[][solid right-pointing triangles]{NeufeldWolfire2017},
\citet[][empty right-pointing triangles]{Luo+2022};
in low-mass dense cores by
\citet[][solid circles]{Caselli+1998},
\citet[][empty circle]{MaretBergin2007}, 
\citet[][empty pentagon]{Fuente+2016},
\citet[][empty left-pointing triangle]{Redaelli+2021}
\citet[][empty hexagons]{Bialy+2022};
in high-mass star-forming regions by
\citet[][stars]{Sabatini+2020},
\citet[][solid diamonds]{deBoisanger+1996},
\citet[][empty diamonds]{VanderTak+2000},
\citet[][empty thin diamonds]{Hezareh+2008}, 
\citet[][solid thin diamonds]{MoralesOrtiz+2014};
in circumstellar discs by 
\citet[][solid squares]{Ceccarelli+2004};
in massive hot cores by
\citet[][empty squares]{BargerGarrod2020};
in molecular clouds close to supernova remnants by
\citet[][solid cross]{Ceccarelli+2011},
\citet[][solid plus signs]{Vaupre+2014};
in a protostellar cluster by
\citet[][empty cross]{Ceccarelli+2014};
towards the Galactic centre by
\citet[][$\times$ sign]{Rivilla+2022};
towards the central molecular zone of NGC 253 by
\citet[][empty up-pointing triangles]{Holdship+2022} and
\citet[][solid up-pointing triangles]{Behrens+2022}.
}
\label{Padovani:fig:zvsN}
      \end{minipage}
\end{figure}

Figure \ref{Padovani:fig:zvsN} shows the most up-to-date collection of observational estimates of $\zzeta$ as a function of H$_2$ column density. 
The black solid lines show the trends predicted by theoretical models \citep{Padovani+2018b,Padovani+2022}. The model $\mathscr{L}$, based on Voyager data \citep{Cummings+2016,Stone+2019}, clearly underestimates the ionisation rate in diffuse clouds ($\nhh\lesssim10^{21}~{\rm cm}^{-2}$). 
Even though the Voyager probes have in all probability passed the heliopause, they are far from measuring the intensity of the average Galactic CR flux. 
Therefore, theoretical models empirically assume the existence of a higher CR flux below about 650~MeV, decreasing the spectral slope, $\alpha$, from 0.1 (Voyager-like spectrum) to $-0.8$ (model $\mathscr{H}$) to $-1.2$.
The models thus succeed in providing an envelope of expected values for $\zzeta$ assuming that the origin of CRs propagating in molecular clouds is Galactic.
We note that estimates of $\zzeta$ under the model $\mathscr{L}$ could be due to uncertainties in the chemical model.

\begin{figure}[htp]
      \begin{minipage}{0.60\textwidth}
	\includegraphics[angle=0.0,width=\textwidth]{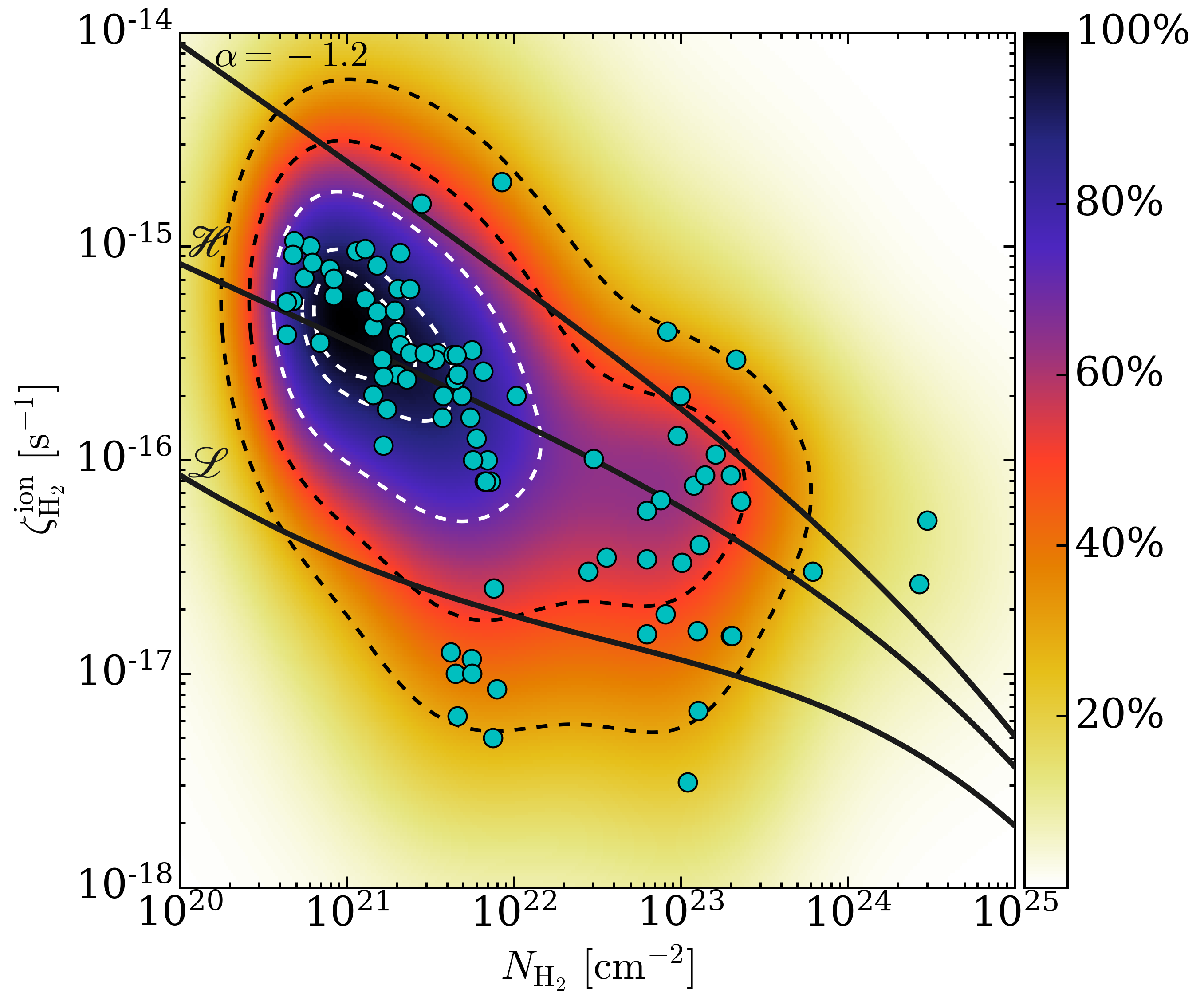}
      \end{minipage}
      \hspace{0.5cm}
      \begin{minipage}[!t]{0.35\textwidth}
	\caption{Kernel density estimation (KDE) of the CR ionisation rate (colour map) 
	by considering only sources ionised by Galactic CRs (see Fig.~\ref{Padovani:fig:zvsN}
	for references to models and observations). Dashed contours show 30, 50, 70, 90, and 95\% of the KDE.
	Solid black lines represent the theoretical models from \citet{Padovani+2022}.
	}
	\label{Padovani:fig:kde}
      \end{minipage}
\end{figure}

The apparent decrease of $\zzeta$ with increasing $\nhh$ is confirmed by the kernel density estimation (KDE) shown in Fig.~\ref{Padovani:fig:kde}. The latter was obtained by considering only $\zzeta$ estimates compatible with the average Galactic CR spectrum, associating a 30\% error with the measurements where the error was not estimated, and removing all upper limits. The KDE contains a sort of bias as most $\zzeta$ estimates are at $\nhh\lesssim10^{22}$~cm$^{-2}$ and around $10^{23}$~cm$^{-2}$, yet the trend is reasonably evident.
This confirms what is predicted by the theoretical models, namely that the flux of CRs (and therefore their ionising power) is attenuated as they propagate in molecular clouds losing energy through collisions with ambient H$_2$.

Figure \ref{Padovani:fig:zvsN} shows a large number of $\zzeta$ that cannot be explained by the average Galactic CR flux. 
These values were measured in molecular clouds near supernova remnants, in protostellar clusters, towards the Galactic centre, and in the central zone of an external galaxy. These recent measurements provide increasingly clear evidence of the presence of locally produced CRs within the sources themselves.
The first models introduced by \citet{Padovani+2015,Padovani+2016} explain the local production of CRs through the acceleration of thermal particles at the shock surfaces (along the protostellar jets and on the protostellar surfaces), according to the first-order Fermi acceleration mechanism
\citep[e.g.][]{Drury+1996}. The maximum energies reached by these CRs are of the order of $0.1-10$~GeV, that is much lower than those expected in supernova remnant shocks. 
However, their flux is sufficient to explain the observations of these extreme $\zzeta$ as well as the synchrotron emission detected in protostellar jets 
\citep[e.g.][]{Sanna+2019} and H\,{\sc ii} regions \citep[e.g.][]{Meng+2019}.
In addition, an important by-product of local acceleration models is the possibility of constraining physical quantities such as magnetic field strength, volume 
density, and flow velocity in the shock reference frame \citep{Padovani+2019,Padovani+2021}.

\section{Local cosmic-ray sources: a new line of research}
In recent years, the theory of local production of CRs within molecular clouds has opened up a new area of investigation. 
Several research groups are developing theories and observational applications on the effects of these local CRs from M-dwarfs on 
Earth-like exoplanetary atmospheres \citep[e.g.][]{Tabataba-Vakili+2016}, 
their propagation in T Tauri winds \citep[e.g.][]{Fraschetti+2018}, 
their effect on protoplanetary discs \citep[e.g.][]{Rodgers-Lee+2017}, 
and their impact in protostellar clusters \citep[e.g.][]{GachesOffner2018,Gaches+2019}.
Research in this new domain will be supported by current and next generation telescopes, such as: 
SKA and its precursors (e.g. MeerKAT, LOFAR), for synchrotron emission at different scales; 
CTA, for $\gamma$-ray emission from high-mass protostars and H\,{\sc ii} regions; 
and ELT/HIRES, JWST, and ARIEL, for the impact of CRs on habitability and exoplanetary atmospheres.

%All standard LaTeX tools are available within the body of the text. Please use 
%\texttt{\textbackslash{}label, \textbackslash{}ref, \textbackslash{}cite} for references and links.
%Because this will be part of a larger document, we request special attention to creating \texttt{\textbackslash{}labels} that are {\it unique} to your article, e.g. by introducing your name in the label like
%\texttt{\textbackslash{}label\{Ossenkopf:sec:main\}}. 

%The use of BibTeX for citations is encouraged. We suggest to use \texttt{astron.bst}. In this way 
%you will automatically generate your references in the correct format via BibTEX
%from your bibliographic database. This file is available from
%\\ \texttt{ftp://ftp.loria.fr/pub/ctan/biblio/bibtex/contrib/astron/}
%(or other CTAN sites). As a citation example that is always worth to be re-read
%we refer to the review by \cite{HollenbachTielens1999}.

\end{document}